# Hydrogen bonding exchange and supramolecular dynamics of monohydroxy alcohols

Shinian Cheng,[1,*] Shalin Patil,[2] and Shiwang Cheng[2,*]

[1]*Department of Chemistry, University of Wisconsin-Madison, Madison, Wisconsin 53706, USA*
[2]*Department of Chemical Engineering and Materials Science, Michigan State University, East Lansing, Michigan 48824, USA*

**Abstract**

This Letter unravels hydrogen bonding dynamics and their relationship with supramolecular relaxations of monohydroxy alcohols (MAs) at intermediate times. Rheological modulus of MAs exhibit Rouse scaling relaxation of $G(t) \sim t^{-1/2}$ switching to $G(t) \sim t^{-1}$ at time $\tau_m$ before their terminal time. Meanwhile, dielectric spectroscopy reveals clear signatures of new supramolecular dynamics matching with $\tau_m$ from rheology. Interestingly, the characteristic time, $\tau_m$, follows an *Arrhenius-like* temperature dependence over exceptionally wide temperatures and agrees well with the hydrogen bonding exchange time from nuclear magnetic resonance measurements. These observations demonstrate the presence of collective Rouse-like sub-chain motions and the active chain-swapping of MAs at intermediate times. Moreover, detailed theoretical analyses point out explicitly that the hydrogen bonding exchange truncates the Rouse-type supramolecular dynamics and triggers the chain-swapping processes, supporting a recently proposed living polymer model.

* Corresponding Authors. scheng235@wisc.edu; chengsh9@msu.edu

Hydrogen bonding (H-bonding) is one of nature's most fundamental interatomic interactions. It is present in alcohols, water, and many bio-macromolecules like proteins, RNA, and DNA, and plays an important role in molecular recognition [1,2], sensing [3], and medication [4-6]. However, a fundamental understanding of the H-bonding interaction and dynamics is far from achieved, especially when H-bonding interaction induces supramolecular structures. For instance, monohydroxy alcohols (MAs), one of the simplest H-bonding liquids, exhibit Debye relaxation [7] from their supramolecular structures, whose origin remains a topic of active debate [8-19]. Recent studies correlate the Debye relaxation with the end-to-end reorientation of MAs with chain structures [11,17,19-21], which initiates speculations of sub-chain collective motions in MAs in analogy to the normal modes of type-A polymers [22]. Although recent studies show signs of supramolecular dynamics at intermediate times in some MAs [21], it is not clear whether or how these dynamics correlate with the sub-chain collective motions. Even less is known about the bridging rules between the reversible H-bonding association/dissociation, *i.e.,* the H-bonding exchange dynamics, and the supramolecular dynamics. Since the H-bonding exchange is much faster than the terminal relaxation of MAs [10,18,21,23], the supramolecular dynamics at intermediate times should serve as a key to unravel the relationship between H-bonding dynamics and supramolecular dynamics of MAs.

Recently, a *zero-free-parameter* living polymer model (LPM) has been proposed to connect the H-bonding dynamics with the supramolecular structure formation and dynamics of MAs [20]. Specifically, the LPM predicts the emergence of an intermediate time associated with the chain breakage, $\tau_B$, that divides the supramolecular dynamics. At time $t < \tau_B$, Rouse dynamics of the supramolecular chains prevail. At $t > \tau_B$, chains swap to speed up the end-to-end vector reorientation leading to Debye relaxation $\tau_D = (\tau_B \tau_\alpha \bar{N}^2)^{1/2}$ with $\tau_\alpha$ being the structural relaxation and $\bar{N}$ the characteristic chain length. Although a scaling of $\tau_D \sim \tau_\alpha^{1/2}$ has been observed recently [20], *no* experiments have been performed to delineate the characteristics of $\tau_B$ and its relationship with $\tau_D$ and H-bonding dynamics, which represent the key differences of LPM from other models, including the transient chain model (TCM) [18], the chain-$g_k$ fluctuation [11] with $g_k$ being the Fröhlich-Kirkwood factor, and the dipole-dipole cross-correlation mechanism (DDCM) [24,25].

In this Letter, we delineate the supramolecular dynamics of MAs with chain structures at intermediate times and elucidate their relationship with the H-bonding dynamics. Different from previous efforts, we rely on new rheological measurements of stress relaxation and relaxation time distribution analyses for dielectric spectra to obtain a complete characterization of supramolecular relaxation without or with active dipolar changes. 2-ethyl-1-hexanol (2E1H), 5-methyl-2-heptanol (5M2H), and 2-butyl-1-octanol (2B1O) have been chosen as model systems due to their known supramolecular chain formation [8]. Poly(propylene glycol) with a number average molecular weight of 4 kg/mol (PPG4k) is included for comparison due to its similar separation between $\tau_\alpha$ and $\tau_f$ with 2E1H (**Figure 1**). Details of materials and methods were presented in **Supplementary Materials (SM)**.

**Figure 1** presents the relaxation modulus, $G(t) = \sigma(t)/\gamma_0$, of 2E1H after a step deformation of $\gamma_0 = 10\%$ at a shear rate of $\dot{\gamma} = 0.5\ s^{-1}$ and $T = 158\ K$. The $G(t)$ of PPG4k at $T = 213\ K$ after $\gamma_0 = 10\%$ is presented for comparison. The time resolution of the stress relaxation experiments is $\sim 0.01s$ that is much smaller than $\tau_\alpha = 2.1 \times 10^{-3}\ s$ of 2E1H and $\tau_\alpha = 3.8 \times 10^{-4}\ s$ of PPG4k at the testing temperatures. Thus, the stress relaxation provides dynamics slower than $\tau_\alpha$, emphasizing their supramolecular nature. For PPG4k, a power-law of $G(t) \sim t^{-1/2}$ is observed followed by an exponential decay with a terminal time, $\tau_f$. The relaxation function can be well described by $G(t) \sim (\tau_f/t)^{1/2} \exp(-t/\tau_f)$ (the solid pink line in **Figure 1**), which is consistent with the Rouse model [26]. On the other hand, 2E1H exhibits dynamics slower than $\tau_\alpha$ due to the supramolecular structures. A two-step relaxation can be resolved that divides the relaxation curve into three regions: Region I, $G(t) \sim t^{-1/2}$ (Rouse scaling) before an intermediate time, $\tau_m^R$; Region II, $G(t) \sim t^{-1}$ between $\tau_m^R$ and $\tau_f$; and Region III, $G(t) \sim \exp(-t/\tau_f)$ at $t > \tau_f$. Experimentally, the single mode terminal relaxation in Region III of 2E1H echoes with the Debye relaxation of 2E1H and $\tau_f \approx \tau_D$ holds [20]. One can also obtain an extrapolated relaxation function for the stress relaxation of 2E1H (see **Section 7** in **SM**):

$$G(t) = A * \left(\frac{\tau_m^R}{t}\right)^{\frac{1}{2}} \exp\left(-\frac{t}{\tau_m^R}\right) + B * \left(\frac{\tau_m^R}{t}\right) \exp\left(-\frac{t}{\tau_f}\right)\left[1 - \exp\left(-\frac{t}{\tau_m^R}\right)\right] \quad (1)$$

with $A$ and $B$ being two fit parameters (the solid blue line in **Figure 1**). The first term on the right of **Eqn. 1** describes the Rouse dynamics of Region I, and the second term covers Regions II and III. Different from PPG4k with the Rouse dynamics dominating the whole stress relaxation, Rouse modes of 2E1H end at $\tau_m^R$ before the terminal relaxation $\tau_f$, highlighting the influence of the reversibility of H-bonding interaction. These observations indicate rich supramolecular dynamics of MAs at intermediate time scales.

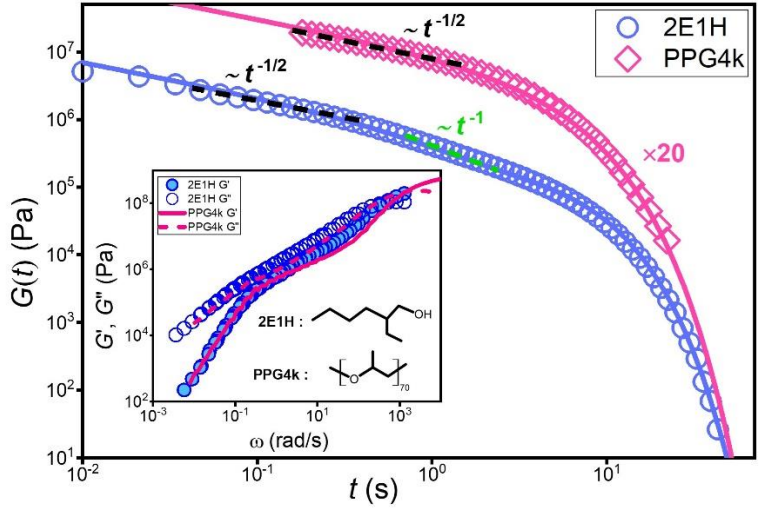

**Figure 1.** The relaxation modulus, $G(t)$, of 2E1H (blue) and PPG4k (pink, 20 times vertical shift) after a step deformation. The blue solid line shows the fit from **Eqn.1**, and the pink solid line represents $G(t) \sim (\tau_f/t)^{1/2} \exp(-t/\tau_f)$. The inset gives the storage modulus, $G'(\omega)$, and loss modulus, $G''(\omega)$, of 2E1H (blue symbols) and PPG4k (pink lines) as well as their molecular structures.

To confirm the revealed Rouse dynamics and to resolve the physical mechanism dictating the crossover between Region I and Region II, we turned to broadband dielectric spectroscopy (BDS). H-bonding-induced supramolecular chains should accompany with an accumulation of dipoles along the chain backbone. If Rouse modes of the supramolecular chains exist, BDS should be able to resolve them [22]. **Figure 2** presents the dielectric loss permittivity, $\varepsilon''(\omega)$ (**Figure 2a**), and derivative spectra of the storage permittivity [27], $\varepsilon'_{der}(\omega) = -\frac{\pi}{2}\frac{\partial \varepsilon'(\omega)}{\partial \ln \omega}$ (**Figure 2b**), at $T = 171\,K$, where $\omega$ is the angular frequency. The corresponding storage permittivity, $\varepsilon'(\omega)$, is presented in **Figure S1** of **SM**. Two characteristic peaks can be resolved from $\varepsilon''(\omega)$ or $\varepsilon'_{der}(\omega)$ with the one at the higher frequencies being the structural relaxation and the lower frequencies the Debye relaxation. We fit the spectra with one Debye function (dashed dot blue) and one Havriliak-Negami (HN) function (dashed green). The solid red lines are the sum of these two functions. A clear short has been observed at intermediate frequencies between the sum and the experiments in both $\varepsilon''(\omega)$ and $\varepsilon'_{der}(\omega)$, especially in $\varepsilon'_{der}(\omega)$. We note that a recent study [28] also commented on the challenge to fit the spectra of other MAs at the intermediate frequencies, which is in agreement with our observations. These observations thus suggest the presence of new relaxation processes between $\tau_\alpha$ and $\tau_D$.

To better resolve the new relaxation processes, we have further performed relaxation time distribution analyses [29,30] that can resolve weak or overlapping processes (see **Section 1.3** in **SM**). This is especially effective for MAs with strong Debye relaxation. **Figure 2c** presents the relaxation time distribution, $\Delta\varepsilon * g(ln\tau)$ with $\Delta\varepsilon$ being the dielectric amplitude and $g(ln\tau)$ the time distribution function, of 2E1H. An intermediate relaxation process is clearly resolved between $\tau_\alpha$ and $\tau_D$, confirming the new supramolecular dynamics of 2E1H at intermediate times not anticipated by the prevailing TCM [18], chain-$g_k$ fluctuation mechanism [11], and DDCM [24,25]. In the following, we focus on the revealed supramolecular dynamics and their relationship with the H-bonding dynamics, which also distinguishes the current contribution from previous works [11,12,15,17,20]. We rely primarily on the relaxation time distribution analyses to identify the intermediate processes to avoid large errors from the fit from introducing another HN function [28].

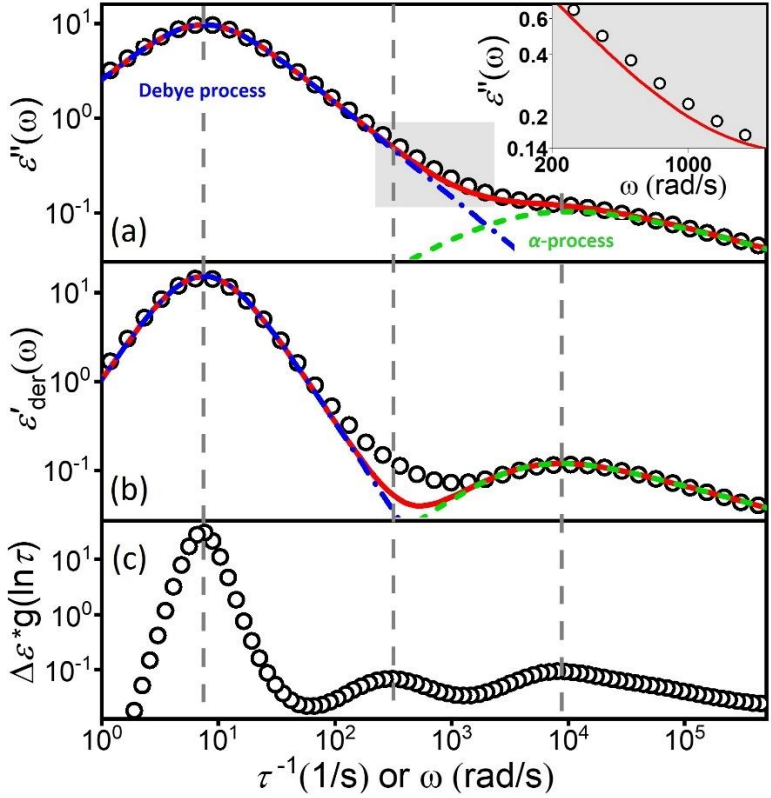

**Figure 2.** (a) The dielectric loss permittivity, $\varepsilon''(\omega)$, and (b) the derivative spectra, $\varepsilon'_{der}(\omega)$, of 2E1H. (c) The relaxation time distribution $\Delta\varepsilon * g(ln\tau)$ of 2E1H. All spectra and analyses were at $T = 171\ K$. The inset of panel (a) shows the amplification of the highlighted region of $\varepsilon''(\omega)$.

**Figure 3** plots a comparison of characteristic relaxation times of 2E1H revealed by rheology and BDS: the structural relaxation, $\tau_\alpha^R$ and $\tau_\alpha^{BDS}$, the intermediate relaxation, $\tau_m^R$ and $\tau_m^{BDS}$, and the Debye and the terminal time, $\tau_D$ and $\tau_f$. Here, the superscript $R$ represents rheology and $BDS$ denotes dielectric measurements. $\tau_\alpha^{BDS}$ and $\tau_\alpha^R$, as well as $\tau_D$ and $\tau_f$, follow identical temperature dependence although $\tau_\alpha^R$ (or $\tau_f$) is ~2.7 times faster than $\tau_\alpha^{BDS}$ (or $\tau_D$). The rheology relaxation time could be slightly different from the BDS measurements owing to the different techniques probing the same physical processes [31]. Interestingly, similar temperature dependence is observed between $\tau_m^{BDS}$ and $\tau_m^R$ at their overlapping temperatures and $\tau_m^{BDS} \approx \tau_m^R$ holds. These results suggest a similar physical origin of the $\tau_m^{BDS}$ and $\tau_m^R$: the Rouse motions of supramolecular chains. In the following, we use $\tau_m$ to represent $\tau_m^{BDS}$ or $\tau_m^R$. Interestingly, detailed analyses of $\tau_m$ show an *Arrhenius-like* temperature dependence over a wide temperature range and times ($10^{-7}$ s to $10^{-2}$ s), where both $\tau_\alpha$ and $\tau_D$ (or $\tau_f$) follow super-Arrhenius temperature dependence. A noticeable deviation from the Arrhenius temperature dependence is observed for $\tau_m$ at temperatures close $T_g$, which will be discussed later.

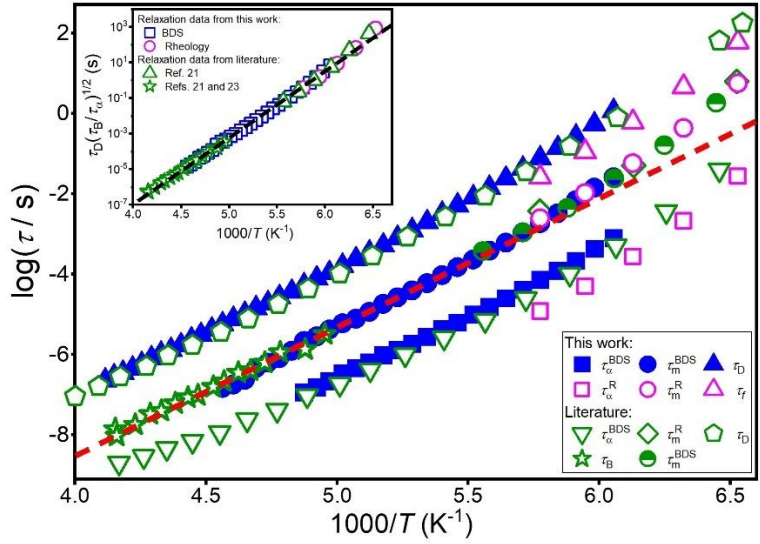

**Figure 3.** Relaxation times of 2E1H from dielectric spectroscopy, rheology, and NMR. The blue filled and pink open symbols are dielectric and rheology results of the current study. The green symbols are literature results of $\tau_\alpha^{BDS}$ [21], $\tau_m^R$ [21], $\tau_m^{BDS}$ [21], $\tau_B$ (from NMR) [23], and $\tau_D$ [21]. The dashed red line shows the Arrhenius-like temperature dependence of $\tau_B$ and $\tau_m$ over wide temperatures. Inset: $\tau_D(\tau_B/\tau_\alpha)^{1/2}$ following an Arrhenius temperature dependence over the whole temperature range. The relaxation data ($\tau_D$, $\tau_B$, and $\tau_\alpha$) to calculate the green triangles and stars are from literature.

To the best of our knowledge, the above experiments and analyses thus provide the *first* direct support for the presence of sub-chain Rouse dynamics of MAs that do not exist in the TCM [18], the chain-$g_K$ fluctuation [11], and the DDCM [24,25]. On the other hand, a recently proposed LPM [20] does predict the emergence of Rouse dynamics up to the chain breakage at $\tau_B$. Specifically, the LPM predicts conventional polymer dynamics at $\tau_\alpha < t < \tau_B$ when chain breakage has not yet taken place (**Figures 4a-4b**). For unentangled polymers, Rouse modes control dynamics slower than $\tau_\alpha$, and a relaxation modulus scaling $G(t) \sim t^{-1/2}$ is anticipated. If no chain breakage takes place before the longest Rouse time of the unentangled polymer, $\tau_B \gg \tau_c \approx \tau_\alpha \bar{N}^2$, the end-to-end reorientation of the supramolecular chains will accomplish through classical Rouse dynamics. As long as $\tau_B < \tau_c$, the chain breakage takes place at $\tau_B$, which truncates the Rouse dynamics and facilitates the end-to-end reorientation of MAs (**Figures 4b-4c**).

Therefore, the LPM explains the $G(t) \sim t^{-1/2}$ at intermediate times. However, the LPM further predicts the Rouse scaling ends at $\tau_B$. Thus, one has to compare $\tau_B$ with $\tau_m$ to enable a full examination of the LPM since both rheology and dielectric spectroscopy do *not* provide direct quantification of $\tau_B$. **Figure 3** compares $\tau_B$ of 2E1H from nuclear magnetic resonance (NMR) experiments (green stars) [23] with $\tau_m$ from rheology or dielectric spectroscopy. Remarkably, $\tau_B \approx \tau_m$ is observed at their common temperature region that covers ~1.5 decades in time. Furthermore, $\tau_B$ from NMR measurements follow identical Arrhenius temperature dependence as $\tau_m$. These observations thus suggest the chain breakage through H-bonding exchange truncates the Rouse dynamics (**Figures 4a-4b**) before the terminal relaxation (**Figures 4a-4c**), supporting the LPM.

Both the rheology and dielectric measurements point to (i) an Arrhenius temperature dependence of $\tau_m$ over wide temperatures; (ii) slight deviation of $\tau_m$ from the Arrhenius temperature dependence taking place at temperatures close $T_g$. In addition, the rheological measurements reveal stress relaxation switching from $G(t) \sim t^{-1/2}$ to $G(t) \sim t^{-1}$ at $\tau_m$. Can the LPM explain these new observations that have not been anticipated by any other existent theories and models? According to LPM [20],

$$\tau_m \approx \tau_B \approx \frac{1}{k_2 \bar{N}} \approx \sqrt{\frac{2}{c_0 k_1 k_2}} \quad (2)$$

with $k_1$ and $k_2$ being the reaction rate constants of H-bonding association and dissociation, $c_0$ is the molar concentration of MAs in the supramolecular chains, and $\bar{N} = \sqrt{\frac{c_0 k_1}{2 k_2}}$ being the characteristic supramolecular chain size [20]. Since $k_1$ and $k_2$ follow Arrhenius temperature dependence, **Eqn. 2** gives an Arrhenius temperature dependence of $\tau_B$ when the temperature dependence of $c_0$ is small. At low temperatures, $c_0$ might change due to the polymerization with reversible bonds [32,33], leading to a deviation of $\tau_B$ from Arrhenius temperature dependence. Indeed, a nice Arrhenius temperature dependence is obtained at the whole temperature range if one plot $\tau_B \bar{N} \equiv \frac{1}{k_2} = \tau_D (\tau_B/\tau_\alpha)^{1/2}$ bypassing the $\bar{N}$ (or $c_0$) in the analyses (**Figure 3 inset**). We note that $\frac{1}{k_2}$ represents the lifetime of H-bonding, which is challenging to obtain in the deep supercooled region [34-36]. Therefore, LPM explains well the Arrhenius temperature dependence of $\tau_m$ as well as its deviation from the Arrhenius temperature dependence close $T_g$.

Regarding the switch in stress relaxation from $G(t) \sim t^{-1/2}$ to $G(t) \sim t^{-1}$ at $\tau_m$, a first principle derivation is not available at this moment that requires a detailed description of the chain-swapping process. However, the LPM predicts $\tau_B = \tau_m \approx \tau_\alpha \bar{N}_R^2$ with $\bar{N}_R$ being the characteristic sizes of the sub-chain segments enjoying full relaxation at $\tau_B$ (**Figures 4a-4b**). According to the Rouse model, the modulus associated with this time scale should be $G_R \sim \frac{1}{\bar{N}_R}$. On the other hand, the full end-to-end vector reorientation relaxation (**Figure 4a-4c**) of the supramolecular chain with length $\bar{N}$ takes place at $\tau_f$ with a modulus $G_f \sim \frac{1}{\bar{N}}$. Thus, $G_R/G_f \sim \bar{N}/\bar{N}_R \approx (\tau_m/\tau_f)^{-1}$. Thus, the LPM anticipates the scaling of $G(t) \sim t^{-1}$ between $\tau_m$ and $\tau_f$ when active chain-swapping takes place. The hold of $\bar{N}/\bar{N}_R \approx (\tau_m/\tau_f)^{-1}$ (or the observation of $G(t) \sim t^{-1}$) gives $\tau_f \approx \tau_m * \bar{N}/\bar{N}_R$, indicating a chain to go through $\bar{N}/\bar{N}_R$ times (on average) breakage to achieve a full end-to-end reorientation. Note that $\tau_f \approx \tau_m * \bar{N}/\bar{N}_R = 1/(k_2\bar{N}_R)$, the end-to-end reorientation of the supramolecular chain can be understood through the effective breakage of sub-chain of sizes $\bar{N}_R$ (**Figures 4b-4c**) as discussed previously [20]. These analyses thus provide a clear illustration of the relationship between the H-bonding lifetime and the supramolecular dynamics of MAs, including the Debye relaxation (**Figures 4a-4c**).

To examine the robustness of the obtained understanding, we have further performed rheology and dielectric measurements for two other MAs, *i.e.,* 5M2H and 2B1O. The stress relaxation of 5M2H and 2B1O **(Figure S3)** show similar characteristics of two-step stress decline beyond the structural relaxation with $G(t) \sim t^{-1/2}$ switching to $G(t) \sim t^{-1}$ before $\tau_f$. Furthermore, the relaxation time distribution analyses of their dielectric measurements **(Figure S4)** offer clear signatures of the intermediate processes, whose characteristic times agree well with $\tau_m$ from rheological measurements and show Arrhenius-like temperature dependence over wide temperatures **(Figure S5)**. These observations thus demonstrate the *universality* of the revealed Rouse-like supramolecular dynamics and the H-bonding exchange mediated chain-swapping processes before the end-to-end reorientation.

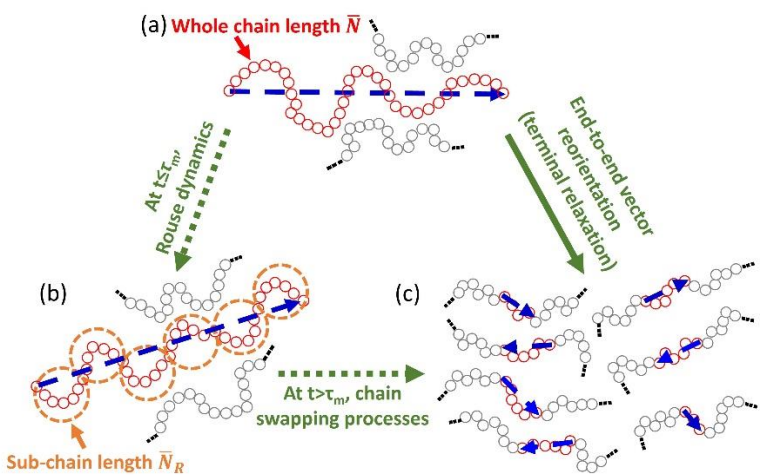

**Figure 4**. Relaxation mechanism of MAs with chain structures, where the grey and red circles represent individual alcohol molecules. The chains with red and grey circles are the test and surrounding chains. The dashed blue arrow represent the end-to-end dipole of the test chain or sub-chain. (a) A representative test chain with length $\bar{N}$ and end-to-end vector at $t = 0$. (b) At $t \leq \tau_m$, chain breakage has not yet taken place and Rouse dynamics dominate. A sub-chain with size $\sqrt{t/\tau_\alpha}$ accomplishes its relaxation at time $t$, and $\bar{N}_R \approx \sqrt{\tau_m/\tau_\alpha}$ is the characteristic sub-chain size that finishes its end-to-end reorientation at $\tau_m$. (c) At $t > \tau_m$, chain breakage must happen. Chain-swapping leads to fragmentation of the original chain (the red circles) and facilitates chain end-to-end reorientation.

In conclusion, we have performed novel rheological measurements and new analyses for dielectric spectra to delineate supramolecular dynamics of MAs with chain structures. Rheological measurements of MAs exhibit Rouse scaling of $G(t) \sim t^{-1/2}$ that changes to $G(t) \sim t^{-1}$ at time $\tau_m$ before their terminal relaxation. At the same time, dielectric analyses reveal new relaxation processes between $\tau_\alpha$ and $\tau_D$ with characteristic time comparable with $\tau_m$ from rheology. Detailed analyses show $\tau_m$ have near *Arrhenius* temperature dependence at high and intermediate temperatures, while slight deviations from the Arrhenius temperature dependence have been observed close to $T_g$. Moreover, $\tau_m$ from rheology and dielectric measurements agree remarkably well with the H-bonding exchange time from NMR measurements. These observations reveal two new features of the relationship between the H-bonding dynamics and dynamics of MAs: (i) the H-bonding exchange truncates Rouse dynamics of MAs; and (ii) the H-bonding dynamics mediated chain-swapping processes at intermediate times. These results have not also been anticipated by TCM [18], chain-$g_k$ fluctuation [11], and DDCM [24,25], and provide the first direct experimental support to a recently proposed living polymer model (LPM), laying down a solid foundation for further elucidations of the relationship between H-bonding dynamics and supramolecular dynamics of H-bonding liquids, including the Debye relaxation--a puzzle in the field for more than 100 years [8].

**Acknowledgment**

This work was supported by Michigan State University Discretionary Funding Initiative (MSU-DFI).

# SUPPLEMENTARY MATERIALS

# Hydrogen bonding exchange and supramolecular dynamics of monohydroxy alcohols

*Shinian Cheng, Shalin Patil, and Shiwang Cheng*

Author Correspondence should be addressed to Shinian Cheng at scheng235@wisc.edu and Shiwang Cheng at chengsh9@msu.edu

## 1. Materials and Methods.

### 1.1 Materials.

2-ethyl-1-hexanol (2E1H, ≥ 99.6%), 5-methyl-2-heptanol (5M2H, ≥ 98%), 2-butyl-1-octanol (2B1O, ≥ 95%), and poly(propylene glycol) with a number average molecular weight of 4 kg/mol (PPG4k) were all purchased from Sigma-Aldrich. All the alcohols were dried by molecular sieves (Sigma-Aldrich) for at least 24 hours to eliminate the moisture before measurements.

### 1.2 Rheology.

Linear viscoelastic properties of all monohydroxy alcohols (MAs) and the PPG4k were characterized on a stress-controlled rheometer (Anton Paar MCR302) equipped with an environmental chamber (CTD600). The environmental chamber is filled with $N_2$ evaporated from a liquid nitrogen unit (EVU 20). The accuracy of the chamber is ± 0.01 K. Small amplitude oscillatory shear (SAOS) was employed to quantify the dynamic mechanical spectra of MAs and PPG4k on a pair of parallel plates with a diameter of 8 mm. The sample gap is controlled to fall in the range of $\sim 0.8 - 1.1 \, mm$. The strain amplitude, $\gamma_\omega$, was set to around $\gamma_\omega = 0.05\%$ at temperatures close to glass transition temperature $T_g$ to $\gamma_\omega = 5\%$ at high temperatures when the dynamic modulus falls below $10^6$ Pa. This protocol ensures the resolution of both the glassy dynamics and the terminal relaxation in the linear viscoelastic spectra measurements. An angular frequency of 628 rad/s to 0.1 rad/s was applied at each test temperature. A thermal equilibrium of 20 min is applied at each test temperature before the measurements.

Stress relaxation after a step deformation $\gamma_0$ was performed to characterize the fine features of the supramolecular dynamics of MAs and PPG4k on the same geometry of a pair of parallel plates with a diameter of 8 mm. A constant shear rate $\dot{\gamma}$ is applied during the step deformation. We varied the $\gamma_0$ and compare the relaxation modulus curves, where the linear response and deviation from the linear response can be identified. All the stress relaxation curves are in the linear response region.

### 1.3 Broadband dielectric spectroscopy (BDS).

The dielectric measurements of 2E1H, 5M2H, and 2B1O were carried out on a Novocontrol Concept-40 system with an Alpha-A impedance analyzer with a ZGS testing interface, and a Quatro Cryosystem temperature controller. The temperature control system has an accuracy of $\pm 0.1\ K$. The measurements were performed on a disk-shaped sample filled in a hollow Teflon spacer with a thickness of 0.14 mm, inner diameter of 16 mm, and outer diameter of 25 mm. The sample is sandwiched by two gold electrodes with a diameter of 20 mm. For each sample, the BDS measurements were performed over a wide temperature range from around $T_g + 100\ K$ to around $T_g - 20\ K$ under cooling at an interval of $5\ K$ and from $T_g - 20\ K$ to $T_g + 100\ K$ upon heating at an interval of $20\ K$. For each temperature, the testing frequency is fixed from $10^7$ to $10^{-2}$ Hz. Thermal annealing of 20 min was applied before each test to ensure thermal equilibrium.

Two types of analyses have been employed to analyze the dielectric spectra of MAs. In the first analysis, a direct fit to the complex permittivity, $\varepsilon^*(\omega)$, is employed through a combination of a Debye function and a Havriliak-Negami (HN) function:

$$\varepsilon^*(\omega) = \frac{\Delta\varepsilon_D}{1 + i\omega\tau_D} + \frac{\Delta\varepsilon_{HN}}{[1 + (i\omega\tau_{HN})^\beta]^\gamma} + \varepsilon_\infty \quad (S1)$$

where $\Delta\varepsilon_D$ and $\Delta\varepsilon_{HN}$ are the dielectric amplitudes of the Debye relaxation and the structural relaxation, $\tau_D$ and $\tau_{HN}$ the HN time of the Debye process and the structural relaxation process, $\beta$ and $\gamma$ the shape parameters, $i = \sqrt{-1}$ the imaginary unit, $\omega$ the angular frequency, and $\varepsilon_0$ and $\varepsilon_\infty$ the vacuum permittivity and the dielectric permittivity at infinity high frequency. Note that the contribution of the dc-conductivity to the relaxation processes is negligible and is omitted in the analysis due to the wide separation between the conductivity relaxation and the Debye process. The characteristic relaxation time of the Debye relaxation and the structural relaxation is obtained from the corresponding HN time. As discussed in the main context, **Eqn. S1** falls short to fit the experiments in the intermediate time scales.

A second method is the relaxation time distribution analysis of the dielectric spectra [1,2]. Specifically, the dielectric relaxation process can be described by a superposition of Debye functions with different relaxation times, $\tau$:

$$\varepsilon^*(\omega) = \varepsilon_\infty + \Delta\varepsilon \int \frac{g(ln\tau)}{1 + i\omega\tau} d\, ln\tau \quad (S2)$$

where $g(ln\tau)$ is the distribution of relaxation time normalized as $\int g(ln\tau)d\, ln\tau = 1$. We applied the generalized regularization method [3] to calculate the relaxation time distribution. It is worth noting that the relaxation time distribution analyses contain the same information on the molecular dynamics as the original dielectric dispersion, although it has the advantage to separate overlapping relaxation processes.

## 2. Dielectric storage spectra, $\varepsilon'(\omega)$, of 2E1H at $T$=171K.

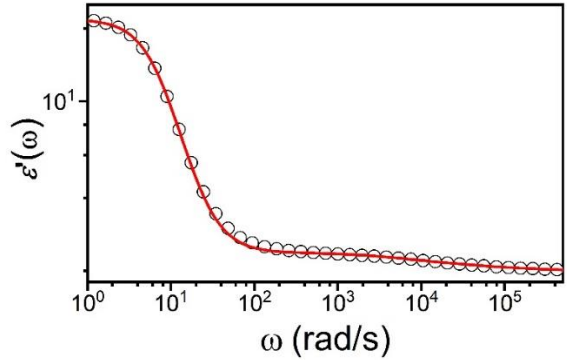

**Figure S1.** Dielectric storage spectra, $\varepsilon'(\omega)$, of 2E1H at $T$=171K (open circles). The red solid line represents the fit using a combination of a Debye function and a HN function.

## 3. Rheology spectra of 5M2H and 2B1O.

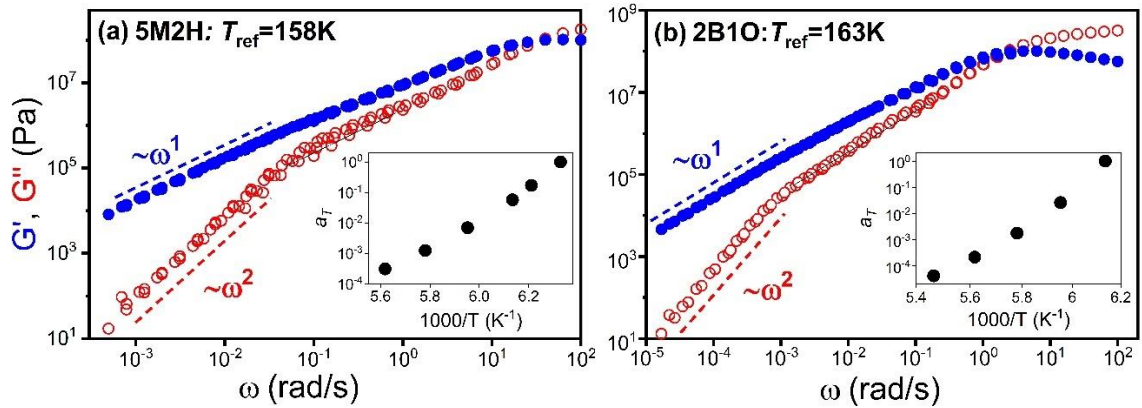

**Figure S2.** Linear viscoelastic master curves of the storage modulus, $G'(\omega)$ (filled blue symbols), and the loss modulus, $G''(\omega)$ (open red symbols), of 5M2H at $T_{\text{ref}}$=158K (a) and 2B1O at $T_{\text{ref}}$=163K (b). The master curves were constructed by assuming the time-temperature superposition. The insets display the corresponding temperature dependence of shift factors, $a_{\text{T}}$, used to construct the master curves.

## 4. Stress relaxation of 5M2H and 2B1O.

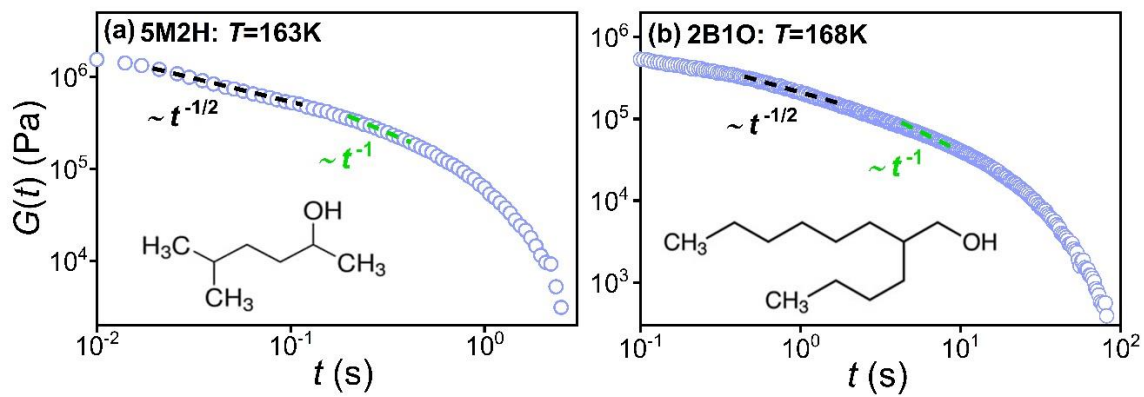

**Figure S3.** The relaxation modulus, $G(t)$, of 5M2H at $T$=163K (a) and 2B1O at $T$=168K (b) after a step deformation. The insets show the molecular formula of these two alcohols.

## 5. Dielectric and relaxation time distribution analyses of 5M2H and 2B1O.

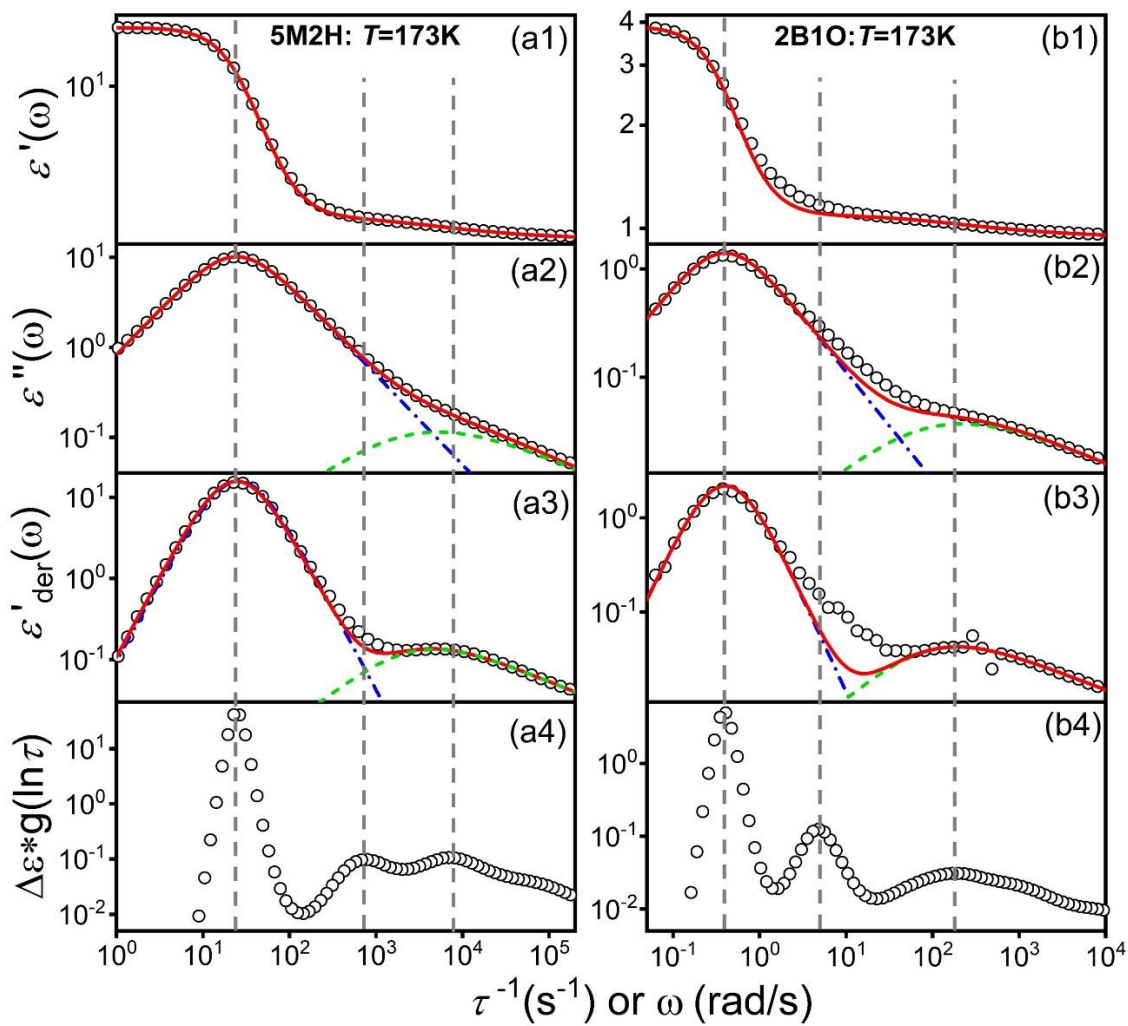

**Figure S4.** The dielectric function and relaxation time distribution of 5M2H at 173K (a) and 2B1O at 173K (b). The solid red lines are the sum of a Debye function (dashed-dot blue) and a Havriliak-Negami function (dashed green). Intermediate processes are clearly visible in both spectra.

## 6. Characteristic relaxation times of 5M2H and 2B1O.

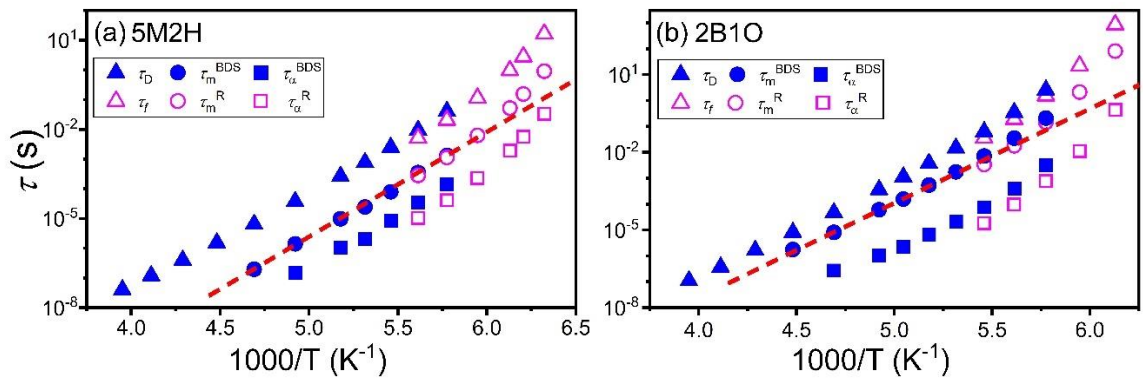

**Figure S5.** Temperature dependence of the characteristic relaxation times of 5M2H (a) and 2B1O (b) from dielectric spectroscopy (blue filled symbols) and rheology (magenta open symbols). The dashed red lines show the Arrhenius temperature dependence of $\tau_m$ at high and intermediate temperatures.

## 7. Derivation of the relaxation function.

The time dependence of the relaxation modulus can be described through[4]

$$G(t) = G_0 \int_0^\infty H(t;\tau) \exp\left(-\frac{t}{\tau}\right) dln\tau \quad (S2)$$

where $G_0$ is a moduli constant, $H(t;\tau)$ is the function of the relaxation time distribution, and $\tau$ is the corresponding relaxation time.

At $t < \tau_m$, the stress relaxation curve gives a scaling of $G(t) \sim t^{-1/2}$ that ends at $t = \tau_m$. Thus, the relaxation time distribution function follows

$$H(t;\tau) \approx \left(\frac{\tau_m}{\tau}\right)^{\frac{1}{2}} \exp\left(-\frac{t}{\tau_m}\right) \quad (S3).$$

This gives

$$G(t) = G_1 \int_0^\infty \left(\frac{\tau_m}{\tau}\right)^{1/2} \exp\left(-\frac{t}{\tau_m}\right) \exp\left(-\frac{t}{\tau}\right) \frac{d\tau}{\tau} = G_1 \sqrt{\pi} \left(\frac{\tau_m}{t}\right)^{\frac{1}{2}} \exp\left(-\frac{t}{\tau_m}\right) \quad (S4)$$

where $G_1$ is the corresponding modulus at short time scales.

At $t > \tau_m$, the relaxation modulus follows $G(t) \sim t^{-1}$ until $t = \tau_f$, which indicates

$$H(t;\tau) \approx \frac{\tau_m}{\tau} \exp\left(-\frac{t}{\tau_f}\right) \quad S(5).$$

This leads to

$$G(t) = G_2 \int_{\tau_m}^{\infty} H(\tau) \exp\left(-\frac{t}{\tau}\right) d\ln\tau = G_2 \frac{\tau_m}{t} \exp\left(-\frac{t}{\tau_f}\right)\left[1 - \exp\left(-\frac{t}{\tau_m}\right)\right] \quad (S6)$$

with $G_2$ being the relaxation modulus at $t = \tau_m$.

Combining **Eqn. S4** and **Eqn S6**, the total stress relaxation function can be written down as:

$$G(t) = G_1 \sqrt{\pi} \left(\frac{\tau_m}{t}\right)^{\frac{1}{2}} \exp\left(-\frac{t}{\tau_m}\right) + G_2 \left(\frac{\tau_m}{t}\right) \exp\left(-\frac{t}{\tau_f}\right)\left[1 - \exp\left(-\frac{t}{\tau_m}\right)\right]$$

$$= A \left(\frac{\tau_m}{t}\right)^{\frac{1}{2}} \exp\left(-\frac{t}{\tau_m}\right) + B \left(\frac{\tau_m}{t}\right) \exp\left(-\frac{t}{\tau_f}\right)\left[1 - \exp\left(-\frac{t}{\tau_m}\right)\right] \quad (S7)$$

which is the **Eqn.1** of the main context that describes well the experiments (**Figure 1**).